# A graph-based narrowband matched-field source localization method

Peng Xiao and Jianmin Yang

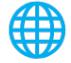

View Online  Export Citation

### ARTICLES YOU MAY BE INTERESTED IN

Data driven source localization using a library of nearby shipping sources of opportunity
JASA Express Letters **1**, 124802 (2021); https://doi.org/10.1121/10.0009083

Physics-informed neural networks for one-dimensional sound field predictions with parameterized sources and impedance boundaries
JASA Express Letters **1**, 122402 (2021); https://doi.org/10.1121/10.0009057

Discrete constriction locations describe a comprehensive range of vocal tract shapes in the Maeda model
JASA Express Letters **1**, 124402 (2021); https://doi.org/10.1121/10.0009058

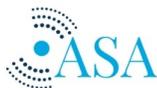

**Advance your science and career as a member of the ACOUSTICAL SOCIETY OF AMERICA**

LEARN MORE

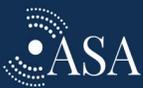

# A graph-based narrowband matched-field source localization method

Peng Xiao and Jianmin Yang[a]

School of Marine Engineering and Technology, Sun Yat-sen University, Zhuhai, Guangdong, 519000, China

xiaop36@mail.sysu.edu.cn, yangjm33@mail.sysu.edu.cn

**Abstract:** Matched field processing (MFP) has been regarded as one of the most successful acoustical methods for positioning underwater sources. In this paper, the narrowband MFP method is combined with a recently developed framework—the graph signal processing (GSP) method. Following the paradigm of GSP, a spatial adjacency matrix is constructed for the arbitrary distributed sensors based on the Green's function, then the source is located by utilizing the graph Fourier transform. The simulation results illustrate that the graph-based MFP outperforms the conventional MFP processors for its better accuracy and fewer requirements for sensor numbers. © 2021 Author(s). All article content, except where otherwise noted, is licensed under a Creative Commons Attribution (CC BY) license (http://creativecommons.org/licenses/by/4.0/).

## 1. Introduction

Matched field processing (MFP) has been regarded as one of the most successful acoustical methods for positioning underwater sources (Baggeroer and Kuperman, 1993; Bucker, 1976; Michalopoulou et al., 2021; Zhang et al., 2021). Generally speaking, MFP can be seen as a beamformer, which matches the measured data with the dictionary of replicas to locate the sources. Since MFP utilizes prior information of propagating channels, it achieves better performance than the traditional geometry locating methods, especially for long-range problems. However, MFP also suffers from shortages such as the heavy computational stress and the sensitivity to the model mismatch (Mantzel et al., 2012).

Currently, the graph signal processing (GSP) method is attracting an explosion of interest (Sandryhaila and Moura, 2014). GSP was developed as a tool for processing large-scale data on irregular domains, including social networks, transportation networks, power grids, immunization and epidemiology networks, and many more. Interestingly, the analogous connections between GSP and classical signal processing could help one to understand the core ideas of GSP (Sandryhaila and Moura, 2013)—such as graph filters, graph Fourier transform, graph frequency, spectral decomposition, etc. GSP is also well fitted in modeling and analyzing the complex data and interactions on the sensor arrays (Xiao et al., 2020). For example, by designing an appropriate graph, or an adjacency matrix, that reveals the intrinsic data structure, GSP can be employed to estimate the direction of arrival (DOA). Moreira et al. (2019) built a cyclic graph structure in the spatial and temporal domains such that the array signal becomes an eigenvector of the adjacency matrix corresponding to the unit eigenvalue. Proudler et al. (2020) further developed the approach by choosing a fully connected graph and highlighting the performance difference between the GSP and MUSIC methods.

Inspired by previous studies, in this paper we implement the GSP method into the narrowband MFP location model. The adjacency matrix is designed for an arbitrarily distributed array based on the Green's function. We show that this GSP-based MFP processor leads to improved localization performance in comparison to that of a conventional Bartlett processor, providing a more accurate estimation with fewer sensors.

The paper is organized as follows. In Sec. 2, the fundamentals of MFP and GSP are introduced. Then the proposed graph-based MFP method is described in Sec. 3, followed by Sec. 4 which presents the simulation results. At last, the conclusions are summarized in Sec. 5.

## 2. Preliminaries

In this section we will give a brief introduction for the MFP and GSP methods.

### 2.1 MFP

At a fixed point $\mathbf{r}(r,z)$, the acoustic pressure field $p(\gamma_0, \mathbf{r}, \omega)$ excited by a point source located at $\gamma_0(r_0, z_0)$ satisfies the wave equation (Jensen et al., 2011), i.e.,

---

$$\nabla^2 p(\gamma_0, \mathbf{r}, \omega) - \frac{1}{c^2(\mathbf{r})} \frac{\partial^2 p(\gamma_0, \mathbf{r}, \omega)}{\partial t^2} = f(\gamma_0, \omega), \quad (1)$$

where $\omega$ is the signal frequency, $f(\gamma_0, \omega)$ presents the source injection at frequency $\omega$, and $r$ and $z$ stand for the receiving range and depth, respectively. Considering that the narrowband condition is assumed throughout the paper, we omit the frequency $\omega$ in the following derivations. If a $N$-element receiver array is deployed to record the pressure field, the pressure vector is obtained as $\mathbf{p}(\gamma_0, \mathbf{r}) \in \mathbb{C}^{N \times 1}$ with its $n$th element being denoted as $p(\gamma_0, \mathbf{r}_n)$. For an ideal environment, the pressure vector could be precisely modeled as $\mathbf{p}(\gamma_0, \mathbf{r}) = f(\gamma_0)\mathbf{g}(\gamma_0, \mathbf{r})$, where $\mathbf{g}(\gamma_0, \mathbf{r}) = [g(\gamma_0, \mathbf{r}_1), ..., g(\gamma_0, \mathbf{r}_N)]$ presents the vector of Green's function. The Green's function can be computed by various propagation models, such as normal mode theory (Porter and Reiss, 1984), ray theory (Porter and Bucker, 1987), and parabolic equation theory (Collins, 1993). However, in the practical environment, the measured pressure vector is always contaminated by the ambient noise. Hence in the rest of the paper, $\mathbf{p}(\gamma_0, \mathbf{r})$ denotes the ideal pressure vector which is computed by the propagation models, and $\bar{\mathbf{p}}(\gamma_0, \mathbf{r}) = \mathbf{p}(\gamma_0, \mathbf{r}) + \mathbf{N}$ denotes the measured pressure vector which contains the noise $\mathbf{N}$, $\mathbf{N} \in \mathbb{C}^{N \times 1}$. Moreover, the measured covariance matrix of the array data is $\mathbf{\Omega} = \bar{\mathbf{p}}(\gamma_0, \mathbf{r})\bar{\mathbf{p}}^H(\gamma_0, \mathbf{r})$, where $(\cdot)^H$ is the Hermitian transpose.

MFP is the three-dimensional generalization of the conventional lower-dimensional, plane wave beamformer. The generalized beamformer matches the measured field $\bar{\mathbf{p}}(\gamma_0, \mathbf{r})$ with replicas of the expected field $\mathbf{p}(\hat{\gamma}_0, \mathbf{r})$ for a test point source at $\hat{\gamma}_0$ (Bucker, 1976). When the test point source is at the true location, the correlation of the two fields will be a maximum, i.e., the fields are matched. Mathematically, this corresponds to the commonest MFP processor—Bartlett processor (Bucker, 1976), which is expressed as

$$B_{\text{Bart}}(\hat{\gamma}_0) = \mathbf{w}^H(\hat{\gamma}_0, \mathbf{r})\mathbf{\Omega}\mathbf{w}(\hat{\gamma}_0, \mathbf{r}), \quad (2)$$

where $\mathbf{w}(\hat{\gamma}_0, \mathbf{r}) = \mathbf{p}(\hat{\gamma}_0, \mathbf{r})/||\mathbf{p}(\hat{\gamma}_0, \mathbf{r})|| = \mathbf{g}(\hat{\gamma}_0, \mathbf{r})/||\mathbf{g}(\hat{\gamma}_0, \mathbf{r})||$ is the normalized weight vector, and $||\cdot||$ denotes the $\ell_2$ norm. The output of the beamformer $B_{\text{Bart}}$ is referred to as the ambiguity surface, and its peak corresponds to the true location $\gamma_0$.

*2.2 Graph signal processing*

Graph signal processing represents a signal *via* a graph structure $\mathcal{G} = (\mathcal{V}, \mathbf{A})$, where $\mathcal{V} = \{v_0, ..., v_N\}$ is the set of $N$ vertices, and $\mathbf{A}$ is the adjacency matrix representing the dependency or similarity relations between the neighbouring vertices. On the basis, a graph signal is defined as a map that associates a signal sample $\mathbf{s} = [s_1, ..., s_N]^T \in \mathbb{C}^N$ with the vertices $\mathcal{V} = \{v_0, ..., v_N\}$.

To proceed, the definition of graph Fourier transform is given. A classical Fourier transform decomposes a signal into an expansion of the orthogonal operators, and the coefficients of these orthogonal operators are called the frequency spectrum of the signal. The graph Fourier transform is defined in a similar way by using the eigenvectors of $\mathbf{A}$. Assuming $\mathbf{A}$ is diagonalizable, the eigendecomposition of $\mathbf{A}$ can be written as

$$\mathbf{A} = \mathbf{T}\mathbf{\Lambda}\mathbf{T}^{-1}, \quad (3)$$

where the columns $\mathbf{t}_n$ of the matrix $\mathbf{T} = [\mathbf{t}_1, ..., \mathbf{t}_N] \in \mathbb{C}^{N \times N}$ are the eigenvectors of $\mathbf{A}$ and $\mathbf{\Lambda}$ is the diagonal matrix containing the corresponding eigenvalues. The graph Fourier matrix is defined as the inverse of the matrix $\mathbf{T}$, i.e., $\mathbf{F} = \mathbf{T}^{-1}$. Consequently, the graph Fourier transform is denoted as $\tilde{\mathbf{s}} = \mathbf{F}\mathbf{s} = [\tilde{s}_1, ..., \tilde{s}_N]$, with its inverse version

$$\mathbf{s} = \mathbf{F}^{-1}\tilde{\mathbf{s}} = \mathbf{T}\tilde{\mathbf{s}} = \tilde{s}_1\mathbf{t}_1 + \cdots + \tilde{s}_N\mathbf{t}_N. \quad (4)$$

Hence, the eigenvectors of $\mathbf{A}$, columns of $\mathbf{T}$, become the graph spectral components, and the coefficients $\tilde{\mathbf{s}}$ of the decomposition are the spectral coefficients of $\mathbf{s}$. The eigenvalues of diagonal values of matrix $\mathbf{\Lambda}$ are referred to as the graph frequencies.

## 3. Graph-based MFP localization

In this section, we formulate the MFP localization problem on the basis of GSP method. The sensors are assumed to be distributed arbitrarily in the range-depth plane. A proper adjacency matrix, or a graph shift, is designed to describe the structure of the sensor data, then the graph Fourier transform is applied to the sensor data to extract the corresponding eigenvector which contains the information of location.

More specifically, Fig. 1 demonstrates the source-receiver setup in the ocean waveguide. The signal emitted by a source at $\gamma_0$ is received by a $N$-element nearly vertical array. As discussed in Sec. 2.1, ideally, the signal received at the $n$th sensor is $p(\gamma_0, \mathbf{r}_n) = f(\gamma_0)g(\gamma_0, \mathbf{r}_n)$. Therefore, the relationship of the pressure field between two different sensors can be expressed as

$$p(\gamma_0, \mathbf{r}_i) = \beta_{i,j} p(\gamma_0, \mathbf{r}_j), \quad (5)$$

where

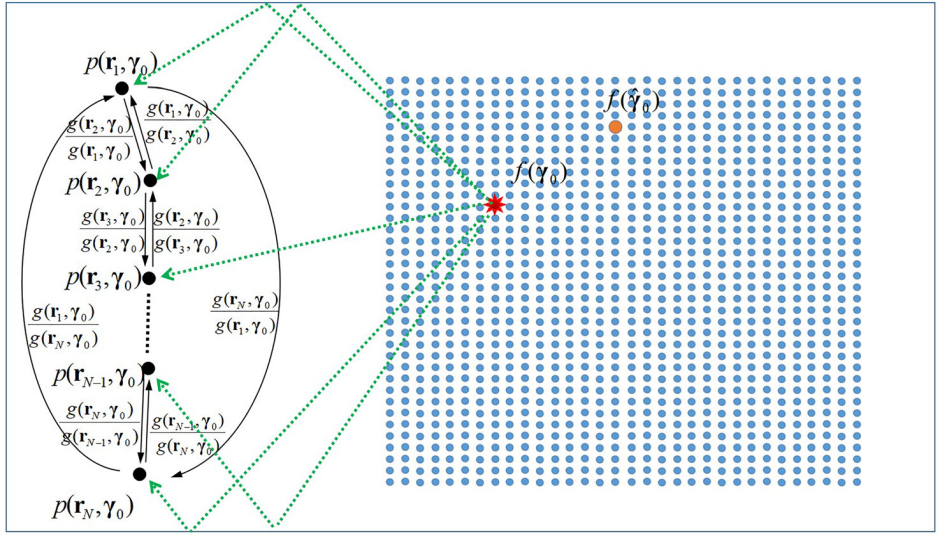

Fig. 1. Schematic plot of the graph-based MFP implementation in the ocean waveguide.

$$\beta_{i,j} = \frac{g(\gamma_0, \mathbf{r}_i)}{g(\gamma_0, \mathbf{r}_j)}. \qquad (6)$$

We can build a space graph shift for the received pressure field upon the relationships of Eqs. (5) and (6). The form of the graph shift matrix is not unique, while in this paper we choose the sparse directed adjacency matrix (Moreira et al., 2019), which connects each sensor with its nearest neighbour (the two sensors at the end also connect with each other). However, it is worth noting that there are other graph matrices, such as the fully connected adjacency matrix which relates any two of the sensors, which can be used in the model. As illustrated in Fig. 1, the sparse directed adjacency matrix is built as

$$\mathbf{A}(\gamma_0, \mathbf{r}) = \frac{1}{2} \begin{bmatrix} 0 & \beta_{1,2} & 0 & \cdots & \beta_{1,N} \\ \beta_{2,1} & 0 & \beta_{2,3} & \ddots & 0 \\ 0 & \beta_{3,2} & \ddots & \ddots & \vdots \\ \vdots & \ddots & \ddots & 0 & \beta_{N-1,N} \\ \beta_{N,1} & 0 & \cdots & \beta_{N,N-1} & 0 \end{bmatrix}. \qquad (7)$$

Like Green's function, the adjacency matrix $\mathbf{A}(\hat{\gamma}_0, \mathbf{r})$ can be generated by the propagation models, and it varies with different source locations. For the correct location $\hat{\gamma}_0 = \gamma_0$, the following graph shift equation holds for the pressure field $\mathbf{p}(\gamma_0, \mathbf{r})$:

$$\mathbf{p}(\gamma_0, \mathbf{r}) = \mathbf{A}(\gamma_0, \mathbf{r})\mathbf{p}(\gamma_0, \mathbf{r}). \qquad (8)$$

In this way, $\mathbf{p}(\gamma_0, \mathbf{r})$ becomes a special eigenvector $\mathbf{t}_u$ of the adjacency matrix $\mathbf{A}(\gamma_0, \mathbf{r})$, which corresponds to the unit eigenvalue. It is worth noting that, compared with (Moreira et al., 2019; Proudler et al., 2020), only a spatial adjacency matrix is used. This is due to the fact that the temporal adjacency matrix requires a precise time shift, which is difficult to satisfy in the real underwater scenarios.

Equation (8) inspires us to apply the graph Fourier transform to the measured data $\bar{\mathbf{p}}(\gamma_0, \mathbf{r})$ to extract the location information. Specifically, the adjacency matrix $\mathbf{A}(\hat{\gamma}_0, \mathbf{r})$ is eigen-decomposed as $\mathbf{A}(\hat{\gamma}_0, \mathbf{r}) = \mathbf{T}\mathbf{\Lambda}\mathbf{T}^{-1}$. When the test point is at the right location, i.e., $\hat{\gamma}_0 = \gamma_0$, the graph Fourier transform $\tilde{\mathbf{p}} = \mathbf{F}\bar{\mathbf{p}}(\gamma_0, \mathbf{r}) = \mathbf{T}^{-1}\bar{\mathbf{p}}(\gamma_0, \mathbf{r})$ presents a one-hot vector with its only nonzero entry corresponding to the eigenvector $\mathbf{t}_u$ (unit eigenvalue). For other cases, i.e., the test point mismatches the right location, the spectral coefficients are not with the one-hot structure. Hence, we design a cost function to evaluate how the test point is matched with the true location,

$$B_G(\hat{\mathbf{r}}) = \left\{ ||[\tilde{\mathbf{p}}]_-||_2^2 \right\}^{-1} = \left\{ ||[\mathbf{T}^{-1}\bar{\mathbf{p}}(\gamma_0, \mathbf{r})]_-||_2^2 \right\}^{-1}, \qquad (9)$$

where $[\mathbf{e}]_-$ denotes deleting the largest absolute value of $\mathbf{e}$. The cost function is slightly different from that used in the previous work of Moreira et al. (2019). Actually, the two functions provide a similar performance while Eq. (9) is simplified such that it is not required to find and delete the eigenvector associated with the unitary eigenvalue before the graph Fourier transform. Hereafter we refer to Eq. (9) as the graph processor. Similar to the conventional MFP processor, the graph processor $B_G(\hat{\mathbf{r}})$ also attains an ambiguity surface, and its peak reveals the correct source location.

## 4. Simulations and discussions

In this section, the performance of the graph MFP processor is evaluated by the numerical simulations, as well as its counterpart—the Bartlett MFP processor.

In the simulations, the locating experiments are conducted within a 100 m deep waveguide, and the sound velocity profile and bottom parameters are shown in Fig. 2. A 10-element vertical line array (VLA) is assumed to be deployed, but the elements are not uniformly distributed. Specifically, we set 41 grids from the surface to the depth of 80 m (grid space $\delta = 2$ m), while each element randomly locates on the grid. The source emitting 200 Hz single-tone signal is set 5 km away from the VLA and 50 m deep from the surface. The two-dimension search area for the source spans from 4 to 6 km in range and from the surface to the bottom in depth. The Green's functions are computed by the standard normal mode code KRAKEN in the frequency domain. It is also assumed that the recorded pressure field $\bar{\mathbf{p}}(\gamma_0, \mathbf{r})$ is contaminated by the Gaussian noise, i.e., $\bar{\mathbf{p}}(\gamma_0, \mathbf{r}) = f(\gamma_0)\mathbf{g}(\gamma_0, \mathbf{r}) + \mathbf{GN}$, where $\mathbf{GN} \sim N(0, \sigma^2 \mathbf{I}) \in \mathbb{C}^N$, and the signal-to-noise ratio SNR is defined as

$$\mathrm{SNR} = 10 \log_{10}\left(\frac{f(\gamma_0)^2 \|\mathbf{g}(\gamma_0, \mathbf{r})\|^2}{N\sigma^2}\right). \tag{10}$$

Throughout the simulations, SNR is set to be 20 dB unless otherwise stated.

Figure 3 shows the ambiguity surfaces for the Bartlett processor (2) and the graph processor (9). Each of the ambiguity surfaces is normalized by its maximum value, and plotted in the decibel scale. It is seen from the figures that the graph processor presents the correct source location, and the corresponding peak on the ambiguity surface is very sharp. The conventional Bartlett ambiguity surface exhibits more lobes than graph processor. However, none of the lobes corresponds to the correct location.

In Fig. 4, it is illustrated for the probability of correct localization versus the number of sensors for the Bartlett processor and the proposed graph processor, and 100 realizations are used to obtain the statistics for each simulation with same number of sensors. Except for the sensor number, the other experimental parameters remain unchanged with Fig. 3. The locating result is considered correct if the estimated depth is within [45 m, 55 m] and the estimated range is within [4.95 km, 5.05 km]. The advantage of the graph processor is clear with a comparison of the two curves: the correct location can be perfectly estimated with ten sensors for the graph processor, while the number needs to be 36 when a conventional Bartlett processor is utilized.

The simulations show the superior performance of the graph processor, and one may be curious about the reason why it works so well. In a previous work (Proudler et al., 2020), the authors find that the graph-based DOA estimation method has some similarities with the MUSIC algorithm (Stoica and Arye, 1989). Hence, we try to understand the mechanism of the graph-based MFP method by a comparison with the MUSIC algorithm. To summarize, the relationships between the graph-based MFP method and the MUSIC method are as follows:

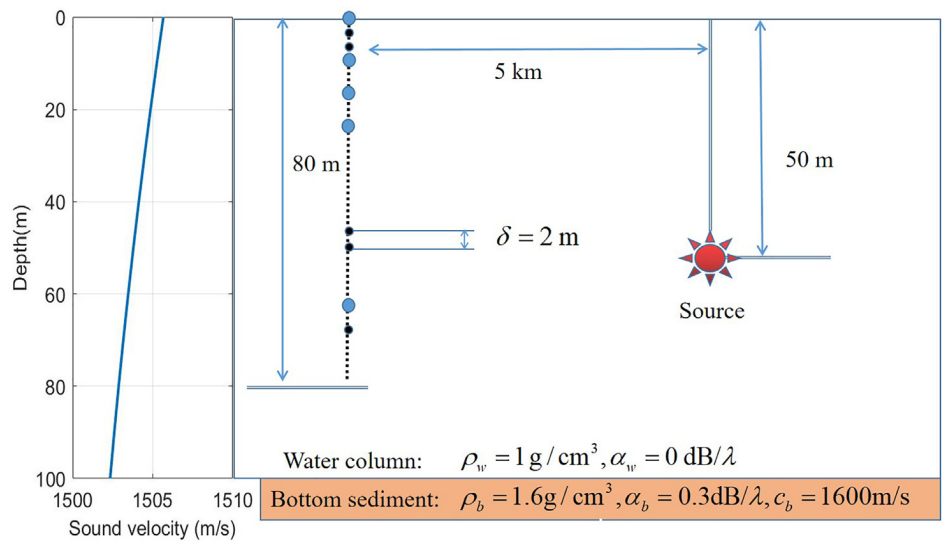

Fig. 2. Simulation environment for the MFP.

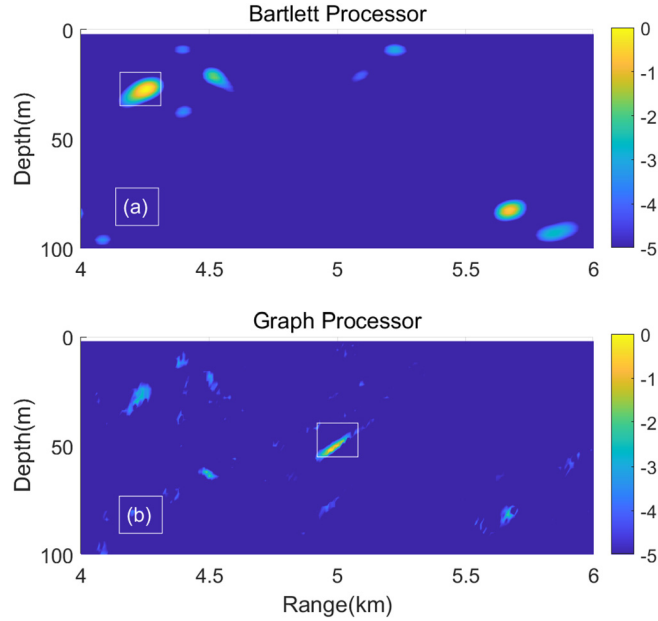

Fig. 3. Localization with two algorithms: (a) Bartlett processor; (b) graph processor.

(1) Both of the algorithms generate the orthogonal basis based on the eigenvector decomposition (EVD). For the MUSIC algorithm, the basis set is obtained by decomposing the covariance matrix; while for the graph MFP algorithm, the basis set is generated from the EVD of adjacency matrix.
(2) Both of the algorithms classify the overall space into subspaces. For the MUSIC algorithm, the overall space is divided into the signal subspace and the noise subspace. Nevertheless, the graph-based MFP processor is suitable for the one source scenario which simply determines the true signal of the source from the others. Hence the graph-based MFP processor divides the space into the true signal subspace and the mismatch signal subspace, and the true signal subspace is spanned by the single eigenvector which corresponds to the unitary eigenvalue.
(3) Both of the algorithms conduct the projection steps. In the graph-based MFP algorithm, the array data is projected onto the orthogonal basis with the graph Fourier transform operation to see how much energy it leaks to the mismatch vectors. For the MUSIC algorithm, the projection step is used to show the energy that signal dimensions leak to the noise subspace.

From the above comparisons, we can see that graph-based MFP method is very similar to the MUSIC algorithm. As the MUSIC algorithm is a high resolution method, we believe the similarities of graph processor to the MUSIC algorithm are the main reasons for the good performance.

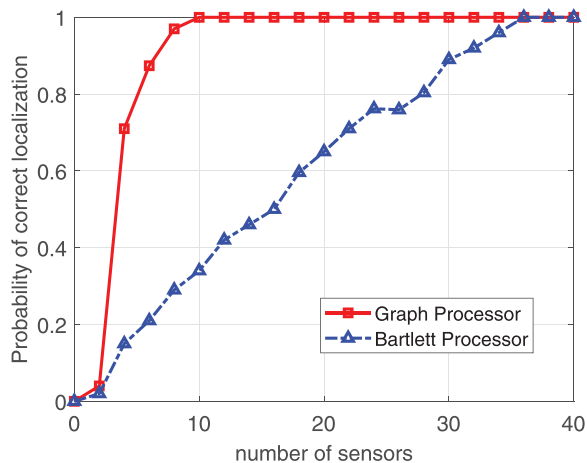

Fig. 4. Probability of correct localization versus the number of sensors for the Bartlett processor and the graph processor.

## 5. Conclusions

In this paper, we propose a new MFP processor combined with the graph signal processing method—the graph processor. The general idea of the processor is to build a graph shift matrix for the replica field on the basis of the computed Green's functions and then to conduct a graph Fourier transform on the receiving data to reveal the source location. The simulations show that, compared with the Bartlett processor, the graph processor can provide a more accurate estimation for the source location, with less sensors. The graph-based MFP processor could be useful in real applications due to its better accuracy and fewer requirements for sensors.

## Acknowledgments

This work was supported by the National Natural Science Foundation of China under Grant No. 61901273.